\documentclass{emulateapj}
\usepackage{amssymb}
\usepackage{epsfig}
\usepackage{natbib}
\usepackage{dsnr}
\shorttitle{}
\shortauthors{Rupke, Kewley, \& Barnes}

\begin{document}

\slugcomment{Accepted by ApJL, 6 Jan 2010}

\title{Galaxy Mergers and the Mass-Metallicity Relation: Evidence for
  Nuclear Metal Dilution and Flattened Gradients from Numerical
  Simulations}

\author{David S. N. Rupke, Lisa J. Kewley, \& Joshua E. Barnes}
\affil{Institute for Astronomy, University of Hawaii, 2680 Woodlawn
  Dr., Honolulu, HI 96822} \email{drupke@gmail.com}

\begin{abstract}

  Recent results comparing interacting galaxies to the
  mass-metallicity relation show that their nuclear oxygen abundances
  are unexpectedly low.  We present analysis of N-body/SPH numerical
  simulations of equal-mass mergers that confirm the hypothesis that
  these underabundances are accounted for by radial inflow of
  low-metallicity gas from the outskirts of the two merging galaxies.
  The underabundances arise between first and second pericenter, and
  the simulated abundance dilution is in good agreement with
  observations.  The simulations further predict that the radial
  metallicity gradients of the disk galaxies flatten shortly after
  first passage, due to radial mixing of gas.  These predictions will
  be tested by future observations of the radial metallicity
  distributions in interacting galaxies.

\end{abstract}


%%%%%%%%%%%%%%%%%%%%%%%%

\section{INTRODUCTION} \label{sec:introduction}

Galaxy interactions are major actors in the process of galaxy
evolution.  Because of this leading role, the impact of interactions
on the chemical evolution of galaxies has become a subject of intense
scrutiny.  We know little about heavy elements in interacting
galaxies, compared to our understanding of more isolated systems.

One place to look for signatures of chemical evolution is the
mass-metallicity relation of galaxies \citep{lequeux79a,rfw84a}.
Low-mass galaxies contain fewer heavy elements than high-mass
galaxies, due to differing star formation histories combined with the
action of gas inflows and outflows \citep{garnett02a,tremonti04a}.  In
particular, star formation enriches the interstellar medium via
nucleosynthesis; gas inflows of relatively metal-poor gas, either from
outside the galaxy or interior to the galaxy, can change the galaxy's
metal distribution; while outflows prevent further enrichment by
blowing out recently produced metals
\citep{edmunds90a,eg95a,ke99a,dalcanton07a}.

Recent studies reveal that galaxies involved in major interactions
fall below this fundamental relation.  In other words, the centers of
interacting galaxies are underabundant compared to galaxies of similar
mass \citep{kewley06b,rvb08a,ellison08a,michel08a,pps09a}.  The
physical model for these deviations from the $M-Z$ relation is
attractively straightforward.  If one or both of the progenitor
galaxies is a spiral galaxy, radial abundance gradients exist in the
merging galaxies \citep{zkh94a,vanzee98a}, with lower abundances at
larger galactocentric radius.  During the merger, low-metallicity gas
from the galaxy outskirts is torqued into the high-metallicity galaxy
center \citep{mh96a,bh96a}, resulting in gas with a lower average
abundance.

This gas redistribution should also result in a change in the gas
metallicity profile across the entire disk.  Thus, radial metallicity
gradients in mergers will differ from those in normal spirals.  Barred
galaxies, which are undergoing radial redistribution of gas due to
bar-induced gas motions, indeed show evidence of either shallower
gradients than are typical of unbarred spirals
\citep{ve92a,zkh94a,dr99a} or flattened oxygen gradients outside the
corotation radius \citep{mr95a,rw97a}.  They also shower lower central
metallicity than galaxies of similar morphological type and luminosity
\citep{dr99a}, as is observed in mergers.  \citet{fbk94a} model this
behavior and demonstrate that the metal gradient changes are due to
gas redistribution by the bar potential.  A brief mention of metal
redistribution in paired galaxies in the context of a coarse-grained
cosmological simulation is also made in \citet{perez06a}.  However,
the idea of radial gas profile changes has not been quantitatively
explored using detailed numerical simulations in the context of
strongly interacting galaxies, which are undergoing even more violent
gas redistributions than barred spirals.

We here analyze N-body / smoothed-particle hydrodynamics (SPH)
simulations of galaxy mergers to quantitatively study metal
redistribution due to merger-induced gas motions.  We wish to explore
nuclear metallicity and radial metal gradients as a function of time
in the context of a merger of equal-mass spirals.  This particular
scenario is relevant to the formation of ultraluminous infrared
galaxies (ULIRGs) and QSOs, which are thought to arise from such a
merger \citep{sanders88a,vks02a}.  \S\ref{sec:simulation-analysis}
introduces our simulations and analysis.  In
\S\S\ref{sec:results}$-$\ref{sec:discussion} we present and discuss
our results, and we summarize in \S\ref{sec:summary}.

\begin{figure*}[ht]
  \plotone{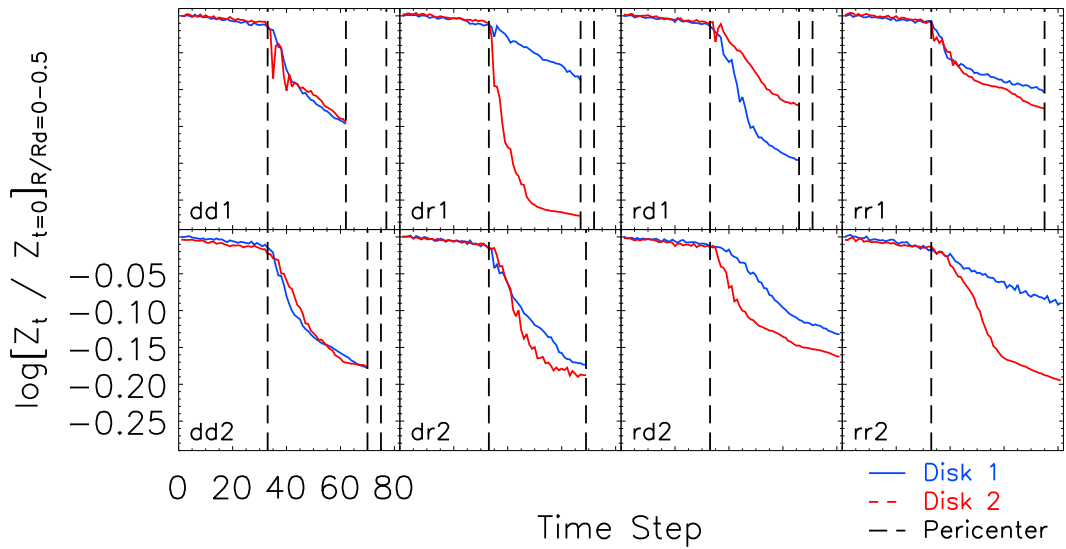}
  \caption{Change in nuclear ($R/R_d < 0.5$) metallicity as a function
    of time.  The metallicity is expressed in terms of $Z$, the mass
    ratio of an element to that of hydrogen.  Strong evolution is
    observed between first and second pericenter, of typical magnitude
    0.2~dex.  The change in nuclear abundance is not strongly
    dependent on geometry, and correlates very well with the amount of
    gas inflow (Figure \ref{fig:npart_v_met}).  Vertical dashed lines
    locate the time of each pericenter, and data beyond the second
    pericenter is not plotted because of intermingling of the two gas
    disks.}
  \label{fig:met_v_time}
\end{figure*}

\section{SIMULATIONS AND ANALYSIS} \label{sec:simulation-analysis}

Our starting point is eight N-body/SPH simulations of close-passage,
equal-mass disk galaxy mergers.  These simulations use the setup
described in in \S3.1 and Appendix A of \citet{barnes04a}, except that
ours exclude star formation.  In brief, 87040 particles (24576 gas)
were used.  Gas mass fractions (as a fraction of the stellar$+$gas
mass) were set at 12.5\%, and the gas was distributed in the same way
as the stellar disk: in an exponential disk with constant scale height
equal to 6\% of the disk scale length.  A cuspy stellar bulge was
included, with a mass 1/3 that of the stellar disk.

The mergers were distributed evenly across the phase space of initial
geometries, with 2 simulations each from direct$+$direct,
direct$+$retrograde, retrograde$+$direct, and retrograde$+$retrograde.
(Direct and retrograde refer to the direction of rotation of each disk
with respect to the encounter orbit.)  For each of these four
combinations, we chose two fairly close encounters with initial
pericenters around $R_{pericenter}/R_{disk} \sim 1$ and 2.  Each
simulation is labeled self-evidently (e.g., dd1 $=$ direct$-$direct
for disks 1 and 2, respectively, and $R_p/R_d \sim 1$).  Exact values
of $R_{peri}$, inclination, and the direction of the line of nodes
were chosen randomly within the constraints of direct/retrograde.

With this set of conditions, we are not probing all of parameter
space, but focusing on the case of a strong interaction between two
present-day spirals of roughly equal mass.  This is most relevant to
the formation of ULIRGs and QSOs in the local universe
(\S\ref{sec:introduction}).  Future work will focus on a wider range
of parameter space.

To get a sense of length scales, we note that for the best-fit
simulation to NGC~4676 considered in \citet{barnes04a}, the linear
scale was such that $R_d = 3$~kpc.  This is comparable to the disk
scale lengths of spirals, which average several kpc \citep{zkh94a}.

We calculate quantities of interest for each disk, assuming that
little mass transfer occurs from disk to disk.  This is a valid
assumption up to the second pericenter, when the disks begin to
overlap appreciably.  We thus constrain our analysis to before second
pericenter.

We include chemical evolution in our analysis by assigning
metallicities to the gas particles at the beginning of the simulation
using the average gradient (per disk scale length) measured by
\citet{zkh94a}, $-$0.2~dex/$R_d$.  The dispersion in the
\citet{zkh94a} sample is 0.1~dex/$R_d$.  We also add to each particle
a normally-distributed random component to the metallicity, with
$\sigma = 0.1$~dex and mean of 0.  We then fix the metallicity of each
gas particle for the duration of the simulation and follow the gas
particles with time.

We neglect enrichment from star formation, since we are focused on
spiral galaxies in the nearby universe.  Gas mass fractions are small
in local, massive spirals ($\sim10$\%; \citealt{kannappan04a}), so
further enrichment will be small compared to that from previous
generations of stars.  (See \S\ref{sec:discussion} for further
discussion.)  We are beginning a study of simulations involving star
formation \citep{sh03a}, which will be considered in future work.

To study the radial metallicity distribution, at each time step we
average azimuthally in radial bins of size 0.1~$R_d$.  We compute the
gradient using a linear fit to the average metallicity in each bin,
weighting each radius equally and fitting from 0 to 5~$R_d$.  (If, as
mentioned above, $R_d \sim 3$~kpc for the progenitor disks, then the
upper radius considered is $\sim$15~kpc).  This is comparable to the
gradient measurement an observer makes, with \ion{H}{2} regions
distributed roughly evenly in radial space.

\begin{figure}
  \plotone{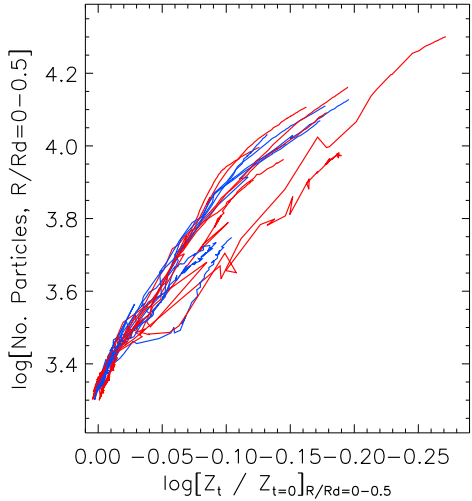}
  \caption{Change in nuclear metallicity vs. number of gas particles
    within the same radius.  Each track represents the time evolution
    of a particular disk (disk 1 in blue, disk 2 in red) for each
    simulation, from lower left to upper right.  The strong
    correlation is unambiguous, demonstrating that the underabundance
    is caused by gas inflow.}
  \label{fig:npart_v_met}
\end{figure}

\section{RESULTS} \label{sec:results}

The finding that motivated this study was the observation that
interacting galaxies have suppressed nuclear metallicities compared to
isolated galaxies of similar luminosity or mass
\citep{kewley06b,ellison08a,rvb08a,michel08a,pps09a}.  In Figure
\ref{fig:met_v_time}, we show that this effect is present in the
simulations.  The metallicity is unchanged prior to first pericenter,
but it drops dramatically after first passage.  By second pericenter,
the change in metallicity ranges from -0.1~dex to -0.3~dex, with an
average of -0.2~dex.  The effect is more dramatic if the initial
gradient is steeper (see below).  The change in nuclear metallicity is
correlated with the amount of gas inflow to the central region (Figure
\ref{fig:npart_v_met}), confirming the suggestion that it is due to
the dilution of the central metallicity as lower-metallicity gas flows
inward from the outskirts of the system.

\begin{figure*}[ht]
  \plotone{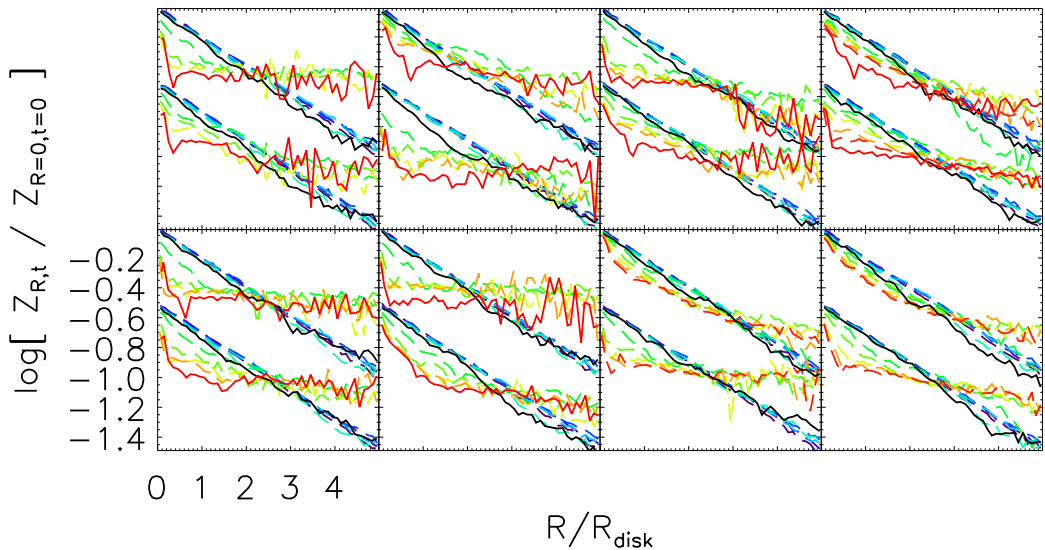}
  \caption{Radial metallicity profile at up to 10 time steps prior to
    second pericenter or the end of the simulation, whichever comes
    first.  Blue (orange) dashed lines indicates early (late) times.
    The profile at first (second) pericenter is shown as a black (red)
    solid line.  The profile for disk 2 is offset by -0.5~dex for
    clarity.  Strong evolution toward a shallow or flat metallicity
    profile is evident between first and second pericenter.}
  \label{fig:gradev}
\end{figure*}

Turning to the metallicity profile of the entire disk, we show in
Figure \ref{fig:gradev} the radial metallicity profile vs. time.
Strong evolution from the initial distribution is evident, with the
disks converging to a shallow or flat distribution over several disk
scale lengths between first and second pericenter.  The profile is not
at all times characterized perfectly by a straight line, and tends
toward a steeper negative slope in the innermost regions.

Radial mixing also results in an increase in the dispersion of the
metallicity of the gas particles.  From an initial dispersion of
0.1~dex, the dispersion within $R/R_d < 0.5$ rises to a value of 0.15
to 0.25~dex at second pericenter.  At larger radii, the change in
dispersion is more dramatic: it jumps to a typical value of 0.2$-$0.25
right after first pericenter.  This dispersion contributes to the
jagged radial profiles at large radius and late times, which is also
caused by small number statistics and discrete tidal structures.

Accompanying the decrease in nuclear metallicity (Figure
\ref{fig:met_v_time}) is an increase in metallicity beyond
$R\sim3~R_d$ (Figure \ref{fig:gradev}).  This results from tidal tails
that have spun out from the interiors of the progenitor disks,
carrying small amounts of more metal-enriched gas from the galaxy
centers into the outskirts.  This removal of metals is insignificant
compared to the effect of radial inflow, in terms of the gas mass
involved.  Thus, the gas removal from the inner regions has no effect
on the central metallicity.  However, because of the relatively small
gas mass at large radii, mixing of small amounts of more enriched gas
easily raises the gas metallicity in these regions.

\begin{figure*}[ht]
  \plotone{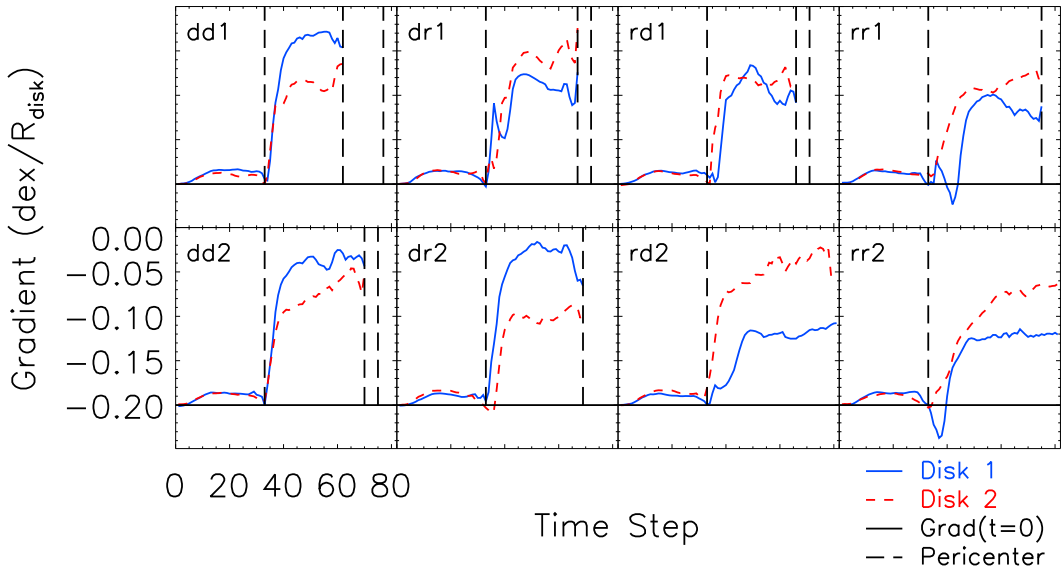}
  \caption{Radial metallicity gradient as a function of time.  As
    suggested by Figure \ref{fig:gradev}, the gradient abruptly turns
    shallow or flattens after first pericenter.  The effect depends
    mildly on geometry, with retrograde$+$retrograde mergers showing
    the least effect.}
  \label{fig:grad_v_time}
\end{figure*}

To quantify the profile evolution, in Figure \ref{fig:grad_v_time} we
show the gradient as a function of time.  We observe the following
about the gradient: (1) it changes little before first passage; (2) it
drops precipitously after first passage (by 0.1 to 0.15
dex/$R_{disk}$); and (3) it experiences mild evolution after this
first drop.  The degree of change varies little with encounter
geometry, but there is evidence that the dd1/2 models experience the
most change, while the rr1/2 models experience the least.

\begin{figure}
  \plotone{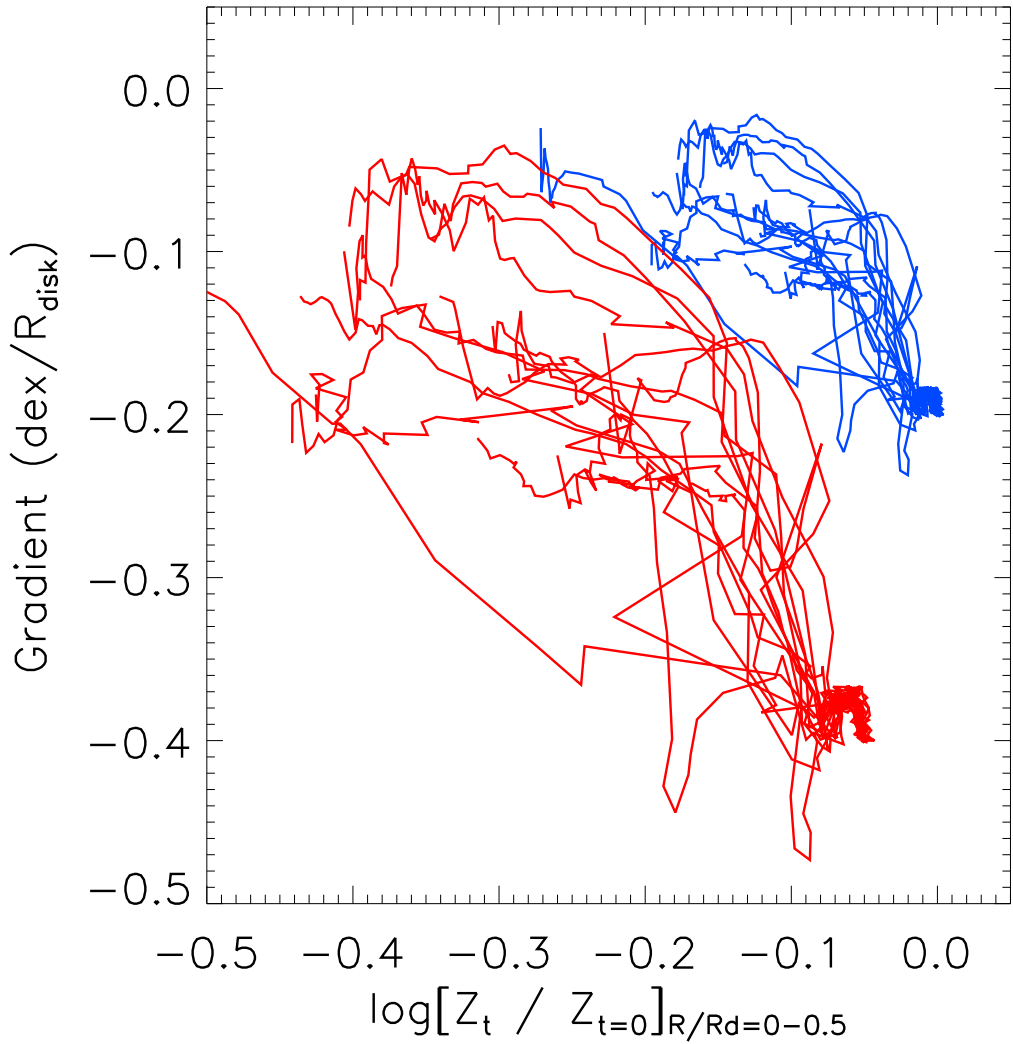}
  \caption{Change in nuclear metallicity vs. radial metallicity
    gradient.  Each track represents the time evolution of a disk,
    from lower right to upper left.  Blue (red) lines represent
    initial gradients of $-0.2$ ($-0.4$) dex/$R_d$.  The two
    quantities are strongly linked, with some scatter.  Steeper
    initial gradients result in lower nuclear metallicities, but still
    converge to shallow or flat gradients.}
  \label{fig:grad_v_met}
\end{figure}

How does the change in gradient relate to the change in nuclear
metallicity, which is more easily measured?  Our models predict that,
in the case of a major merger, the two quantities are loosely
correlated (Figure~\ref{fig:grad_v_met}).  However, the change in
gradient proceeds more quickly than the change in nuclear metallicity,
with the latter being most dramatic after the gradient has already
flattened substantially.  This is due to the large gas mass of the
central region, which requires significant inflow for its metallicity
to be strongly affected.

In Figure~\ref{fig:grad_v_met}, we demonstrate that the flattening of
the gradient is not a strong function of the initial gradient, by
comparing initial gradients of $-0.2$ and $-0.4$~dex/$R_d$, which
correspond to the average and the steepest gradients observed by
\citet{zkh94a}.  The range of gradients near second pericenter is 0 to
$-0.1$~dex and $-0.05$ to $-0.2$~dex, respectively.  However, the
change in nuclear metallicity can be much higher if the initial
gradient is steeper.  More precisely, the final change in nuclear
metallicity (in dex) is roughly equal to the initial gradient (in
dex/$R_d$).

\section{COMPARISON TO OBSERVATIONS} \label{sec:discussion}

Our results on nuclear underabundances compare favorably with
observations.  Previous work on a wide variety of interacting systems
shows that underabundances are typically $0.1-0.3$~dex
\citep{kewley06b,rvb08a,ellison08a,michel08a}, although higher
underabundances exist in some massive, blue systems \citep{pps09a}.
Our simulated offsets are consistent with most of these results,
though an unknown fraction of the observed interacting systems
represent weak or minor mergers.  However, we cannot explain the
strongest deviations of $0.6-0.8$~dex \citep{pps09a} unless the
initial gradients are steeper than those measured by \citet{zkh94a}
{\it or} the progenitors of these interacting systems were already
underenriched due to lack of previous star formation (consistent with
their blue colors).

The present work is most applicable to mergers of equal-mass galaxies;
the only systems that we are sure meet this criteria are ULIRGs.  The
metallicity offsets in these systems average $0.1$~dex and range up to
0.3~dex \citep{rvb08a}\footnote{These numbers are smaller than
  presented in \citet{rvb08a}, and result from an updated analysis
  made possible by a recalibrated mass-metallicity relation using the
  [\ion{N}{2}]/[\ion{O}{2}] metallicity diagnostic \citep{ke08a}.
  This diagnostic is the most robust strong-line diagnostic
  \citep{kd02a}.}.  Again, these results are in excellent agreement
with the simulations, especially if we assume either that the initial
gradients were somewhat shallower than predicted by \citet{zkh94a} or
that there has been modest re-enrichment due to star formation.

There is no published data on the metallicity gradients of interacting
systems.  A study of weakly interacting spirals found no difference in
metallicity gradient compared to isolated spirals \citep{marquez03a},
but the [\ion{N}{2}]/H$\alpha$ metallicity diagnostics used in this
study have significant uncertainties \citep{kd02a}.  We have begun a
program to observe the gradients in strongly interacting galaxies, to
understand how the metal redistribution occurs over the course of a
merger and to compare to simulations.

It is clear that any significant galaxy-galaxy interaction will result
in some change in the metallicity distribution of the galaxies
involved.  These effects may be important when interpreting the
evolution of the mass-metallicity relation.  At higher redshifts, when
the merger and interaction rates were higher than at present
\citep[e.g.,][]{kartaltepe07a}, the dilution of nuclear abundances may
be a more ubiquitous phenomenon.  The effect will be especially
significant if a population chosen to trace the evolution of the
mass-metallicity relation consists of significant numbers of
interacting systems.  For instance, low-redshift Lyman Break Galaxy
analogs show lower than expected metallicity and evidence of a major
merger \citep{hoopes07a,overzier08a}.  LBGs have been used to trace
metallicity evolution at high redshift \citep{erb06a}.  However,
higher redshift galaxies also have higher gas mass fractions, while
our simulations assume a low gas mass fraction.  Further work over a
larger range of initial conditions is needed.

One caveat to our results is the presence of ongoing star formation,
which we have ignored.  Early in the merger, this star formation
occurs across the disk of the galaxy, and hence is unlikely to change
our conclusions about gradients.  Late in the merger, star formation
is heavily concentrated toward the galactic center, and can raise the
heavy element content in these regions.

The magnitude of this effect depends on the initial metallicity and
the fraction of the remaining gas that is consumed in star formation.
Consider two spirals with initial nuclear metallicities of \zsun\ and
a true oxygen yield of 0.005 \citep[][reduced by a factor of two to
account for their high metallicity calibration]{tremonti04a}.  Since a
significant amount of gas remains at the late stages of a merger
\citep{sss91a}, we assume that not more than half of the gas has been
consumed prior to coalescence (more than this, and the initial gas
mass fractions will have been unreasonably high).  In the closed box
model \citep{ta71a}, if 50\% of the total gas mass is consumed in a
solar-metallicity system, the metallicity will rise by at most 70\%,
or 0.2~dex.  In this extreme scenario, enrichment by star formation
could wipe out the nuclear metallicity underabundance predicted by our
simulations, and may affect the gradient as well, depending on where
the star formation occurred.  However, comparison to the data shows
that ongoing enrichment is small up to the ULIRG phase, since
underabundances at the $0.1-0.3$~dex level (or higher) are still
observed in local interacting systems.  This in turn implies that
either we have assumed too much gas consumption in this example
calculation (i.e., that the ``extreme'' case is not valid), or that
outflows (which are ubiquitous in these systems; \citealt{rvs05a})
have expelled metals produced in ongoing star formation, preventing
further enrichment.

Either way, our simulation results appear robust to ongoing star
formation.  However, more sophisticated simulations involving both
star formation and metal redistribution by supernovae and outflows
will be required to confirm this.  We are beginning to analyze
numerical simulations of merging galaxies that include a prescription
for chemical evolution due to star formation and stellar feedback
\citep{sh03a}, especially to understand mergers at higher redshift
with large gas mass fractions.

\section{SUMMARY AND PROSPECTS} \label{sec:summary}

We have presented numerical simulations that confirm that nuclear
metallicity underabundances observed in interacting disk galaxies are
due to merger-driven inflow of low-metallicity gas from the outskirts
of the progenitor disks.  The magnitude of the effect is in agreement
with what is observed ($\sim$0.2~dex;
\citealt{kewley06b,rvb08a,ellison08a,michel08a,pps09a}).  The nuclear
metallicity drop traces closely the amount of merger-driven inflow,
and is accompanied by an increase in the dispersion of gas cloud
metallicities at a given radius.  The metallicity drop and increase in
dispersion occur predominantly between first and second pericenter.

Also evident between first and second pericenter is a dramatic
flattening of the initial radial metallicity gradient.  This
flattening reflects the effects of gas redistribution over the galaxy
disks.  Such gradient changes will be observable in studies of
\ion{H}{2} region metallicities in strongly interacting systems.  By
comparing to detailed simulations, the distributions of gas-phase
metallicities are certain to provide another way to constrain the
evolutionary state of interacting systems.  In particular, they are
very sensitive to the gas motions that occur within the system.

Our results apply primarily to local examples of major mergers of
spiral galaxies.  Analysis of simulations including more sophisticated
chemical evolution prescriptions is currently underway, especially to
probe a wider range of merger initial conditions (including mass ratio
and gas mass fraction).  It is important to understand the effect this
could have on the determination of cosmic metallicity evolution, since
galaxy mergers and interactions were more common at high redshift.
Samples of high redshift galaxies that are dominated by interacting
systems may bias metallicity determinations to low metallicities,
compared to samples at low redshift dominated by isolated systems.
Future work will explore this effect.

\acknowledgments

The authors thank the referee for a helpful report, and Fabio Bresolin
for commenting on the manuscript.  DSNR and LJK are funded by NSF
CAREER grant AST07-48559.

%\bibliography{apj-jour,dsr-refs}

\end{document}